\documentclass[a4paper]{jpconf}
\usepackage{graphicx}
\begin{document}
\title{Uncertainites in AGB evolution and nucleosynthesis}

\author{John Lattanzio$^1$ and Amanda Karakas$^1$}
%The style for the names is initials then surname, with a comma after all but the last two names, which are separated by `and'. Initials should {\it not} have full stops. First names may be used if desired.

%\address{$^1$ Production Editor, \jpcs, \iopp, Dirac House, Temple Back, Bristol BS1~6BE, UK\\
%$^2$ Faculty of Science, The University of Hong Kong, Pokfulam Road, Hong Kong}

\address{$^1$Monash Centre for Astrophysics, School of Phyics and Astronomy, Monash University, 3800, Australia}
\ead{$^1$john.lattanzio@monash.edu}

\begin{abstract}
We summarise the evolution and nucleosynthesis in AGB and Super-AGB stars.
We then examine the major sources of uncertainty, %including a detailed look at the inclusion of 
especiall mass-loss.
\end{abstract}

\newcommand{\Msun}{M_\odot}

\section{Introduction}
AGB stars are those that return to the giant branch after
core helium exhaustion, and show thermal instabilities (or pulses) of the helium
shell while ascending the second giant branch.
They begin their lives with masses between about $0.8$ and $8\Msun$. The Super-AGB stars (hereafter SAGB)
are the higher-mass cohort, from about $8$ to $12\Msun$, which burn carbon in their cores, then proceed to 
show thermal pulses. Their mass range is bounded above by the massive stars, which go on to more
advanced nuclear burning.
The shape of the initial mass function ensures that most stars above $0.8\Msun$ experience an AGB phase. They are
bright, and hence they dominate the light of many galaxies. Their copious mass-loss, and advanced
nucleosynthesis, ensures that they are major producers of dust, possibly producing as much as 90\%
of the dust in our galaxy [1]. Their nucleosynthesis is now understood to be crucial 
for understanding the chemical evolution of the Universe. AGB stars are significant producers of 
carbon, nitrogen, fluorine and about half of the elements heavier than iron.

\section{Summary of AGB evolution}
In what follows we restrict our discussion to single stars.
We will discuss pre-AGB evolution below, when we consider the uncertainties. The structure of an AGB star
is well known.
In the centre is the degenerate core composed mostly of C and O,
surrounded by a He-burning shell. The next layer consists of the ashes of the H-burning shell, 
and is topped by said shell. There
is then usually a small radiative buffer zone, and finally a deep convective envelope.
The AGB phase follows immediately
after the star exhausts its core He supply. While the He shell is becoming established, as a result of the
core contracting and bringing
He at the edge of the core to higher temperatures, there is a substantial energy output and
the outer layers expand and cool. This is called the {\it early AGB\/} and the cooling outer
layers cause convection to extend inwards in mass. For stars exceeding about $3\Msun$ the H shell is extinguished
and the convection may reach into the He-rich region, composed of the ashes of H burning. This is called
the {\it second dredge-up\/} and it can produce substantial increases in the surface He content, up to about 38\% by mass
[2]. It is this event that produces the classic core-envelope structure required for stars to ascend the giant branch, 
and we will see later that the details of the timing and depth of the second dredge-up 
are vital in determining the final fate for SAGB stars [3].

Thermally pulsing evolution has been reviewed previously in the literature, and we refer the reader to [2,4,5] for details. 
Here we give only a summary, due to space limitations. The He shell is thermally unstable and experiences flashes or pulses 
every 10$^3$ to 10$^5$ years, depending on the core mass. During this ``on'' phase the He shell drives a convective region
that extends from the He shell almost to the H shell, and is known as the {\it intershell\/} or {\it pulse-driven\/} 
convective zone. This distributes the products of He burning, mostly C, from the He shell
almost to the bottom of the H shell. This region now comprises about 75\% He and 25\% C (by mass).
After about 200 years the convection ends as the shell energy input decreases.
The regions outside the He shell expand due to the energy input, causing the H shell to be pushed out 
to lower temperatures where it is extinguished (or nearly so). 
The opacity in these cooler regions increases and the bottom of the convective envelope extends inwards in mass.
If the envelope reaches into the (top of the) region that was previously the intershell
convective zone, then newly synthesised C in this region is mixed to the surface. 
This is known as the third dredge-up, and it can happen repeatedly, following (nearly) each 
thermal pulse. Note that a low-mass star may not experience a second dredge-up, but it
may still experience many third dredge-up events. The nomenclature is well established, but not perfect.
After the energy from the thermal pulse has diffused to the surface, the expanded envelope contracts and heats,
and the H discontinuity becomes the new H-burning shell. The star begins the longest phase of the 
pulse cycle, the inter-pulse phase. At this stage the He shell is essentially inert and the star 
is powered by the H shell. This phase ends with the next thermal pulse of the He shell, some $10^3$ to $10^5$ years later,
depending on the (core) mass of the star.

%These phases are shown schematically in Figure~\ref{fig1}.

%\begin{figure}[h]
%\begin{center}
%\includegraphics[width=0.5\textwidth]{AGBslice.jpg}%\hspace{2pc}%
%%\begin{minipage}[b]{0.5\textwidth}
%\caption{\label{fig1}A schematic showing the events that occur in a thermal pulse.}
%%\end{minipage}
%\end{center}
%\end{figure}

\newcommand{\iso}[1]{$^{#1}$}

\section{Summary of AGB nucleosynthesis}
The above explains perhaps the most obvious consequence of pulses and dredge-up, namely the 
increase in the surface C content. But just as important is the synthesis of $s$-process elements,
lithium,  and the
occurrence of hot bottom burning (hereafter HBB). 

%\begin{figure}[h]
%\begin{minipage}{0.4\textwidth}
%\includegraphics[width=0.9\textwidth]{AGBslice.jpg}
%\caption{\label{label2}Figure caption for first of two sided figures.}
%\end{minipage}\hspace{0.02\textwidth}%
%\begin{minipage}{0.58\textwidth}
%\includegraphics[width=0.9\textwidth]{C13.jpg}
%\caption{\label{label3}Figure caption for second of two sided figures.}
%\end{minipage}
%\end{figure}

\subsection{Hot bottom burning}
For the more massive stars, exceeding about $4\Msun$ (for
solar composition, and decreasing as [Fe/H] decreases) the bottom of the convective envelope reaches
into the top of the H-burning shell during the interpulse phase.  Proton captures can occur 
at these high temperatures, and this is called
HBB. To follow it accurately requires simultaneous calculation of mixing and burning, usually done with
a diffusive approximation although it is important to remember that convection is {\it advective} 
rather than {\it diffusive}. The major reactions in such a case are CNO cycles, which
burn \iso{12}C (and at extreme temperatues \iso{16}O) into \iso{13}C and \iso{14}N.  
Often the \iso{12}C is the same \iso{12}C that was dredged-up following the previous thermal pulse. The
result is that HBB can prevent the formation of a C-star, or even reduce the C/O ratio from above unity to below unity
at the cost of increasing the (primary) \iso{14}N content of the envelope. For higher temperatures, 
which means for more massive AGB stars or even SAGB stars, we find the activation 
of the Ne-Na chain and possibly the Mg-Al chain [6,7] or even beyond, for SAGB stars.

\subsection{Lithium production}
Lithium continues to cause headaches for astrophysics.  It is very
fragile and burns through pp chains at temperatures as low as $2.5$\,MK.
The AGB stars contribute to the Li problems through being production sites for Li,
but with an uncertain effect on Galactic Li content.

If the AGB star experiences HBB then it will produce Li through the Cameron-Fowler mechanism
[8,9]. This seems to be verified by observations [10,11], which show that Li is present in 
AGB stars that are not C-stars, as expected when HBB operates. The problem is that the overall 
production or destruction of Li in these stars is unclear because of the uncertainties 
associated with mass loss [12,13]. Although Li can be efficiently produced, it is also
destroyed by proton captures in the hot bottom of the convective envelope. 
The production and destruction is a delicate balance, with production driven by the 
initial supply of \iso{3}He. When that is used up then the destruction will dominate. Hence the 
Li-rich phase is only temporary. Whether the overall yield of Li from such a star is
positive or negative depends on when the majority of the mass-loss occurs. 
If it is after the destruction, as most models predict, then the stars are negligible 
producers of Li for the Galaxy. If, however, the mass-loss rate is high when Li is 
abundant then the stars can produce significant amounts of Li. Li is also a crucial 
guide to extra-mixing processes on the RGB, as we discuss below.

\subsection{Producing s-process elements}
The slow neutron capture process, or $s$-process, is responsible for producing about half of the elements heavier than Fe,
and AGB stars are the main producers [14]. There are two ways for this to occur. The first is 
active in more massive AGB stars, and occurs in the intershell convective region during a thermal pulse.
The H-shell transmutes essentially all CNO species into \iso{14}N. During the flash these \iso{14}N nuclei 
can capture two $\alpha$ particles (He nuclei) to produce \iso{22}Ne. If the temperature exceeds about 300\,MK
then one more $\alpha$ capture can occur, producing free neutrons via \iso{22}Ne($\alpha$,n)\iso{25}Mg and 
these neutrons are now available for capture on Fe and other heavy elements and
%The resulting neutron densities are relatively low, and 
the production follows the $s$-process path.
The elements produced in this way are then mixed to the envelope at the next dredge-up episode.

%\begin{figure}[h]
%\begin{center}
%\includegraphics[width=0.5\textwidth]{C13.jpg}%\hspace{2pc}%
%%\begin{minipage}[b]{0.6\textwidth}
%\caption{\label{C13}A schematic of the activation of the \iso{13}C neutron source during a thermal pulse.}
%%\end{minipage}
%\end{center}
%\end{figure}

The other way to produce neutrons is through the \iso{13}C source. %, and is sketched in Figure~\ref{C13}.
This requires some form of partial mixing of protons below the formal bottom of the convective 
envelope at the end of dredge-up. There is much debate about the mechanism and how exactly this occurs (see discussion in [2]).
But let us suppose it happens, as seems to be required by observations. Then these protons can be captured by
the abundant \iso{12}C to produce \iso{13}C which can then capture an $\alpha$ particle to produce a free 
neutron (and a \iso{16}O nucleus). These neutrons produce $s$-processing as before, but at a lower neutron density 
than the \iso{22}Ne source, albeit for a longer time. We note that the formation and properties of the
\iso{13}C pocket are major uncertainties in our understanding of the $s$-process.

\section{Super-AGB stars}
Recent years have seen a rise in the number of studies of SAGB stars. These are very demanding 
calculations and it is only recently that we have had the computer power to throw at this problem.
SAGB stars ignite C in their cores, whereas the normal AGB stars do not. Subsequent evolution 
depends critically on the mass (and the details of how convection is calculated [15]). 
The C ignites off-centre in a small convective shell. This shell may burn all the C present in the
convective region into Ne, and then a second shell ignites C, and so forth. 
These shells can be located further toward the centre or further outward, with the result that 
one eventually burns all of the core C in some cases, and in others the 
C burning may not reach the centre so we have a CO core and an ONe outer region.
The crucial thing is that these stars then proceed to experience thermal pulses on the (S)AGB.
They are quite separate beasts to ``massive'' stars, which go on to further nuclear burning 
stages after C burning. SAGB stars do not.

SAGB evolution is qualitatively the same as AGB evolution, but the quantitative differences are important.
Firstly, the HBB occurs at high temperature [6,7,16] due to the deep convective envelopes. However, the
intershell convective zone is very small in mass, typically $10^{-3}$ to $10^{-5}\Msun$ (compared to something like
$0.01\Msun$ for AGB stars). This means that the region undergoing neutron captures (from the \iso{22}Ne source)
is tiny and when this region is diluted in the much deeper envelope (the core mass is perhaps $1\Msun$, 
leaving an envelope mass of a few $\Msun$) then the enhancements of $s$-process elements are not expected to be large.
This may change near the end of the evolution, when the envelope mass has been dramatically reduced due to mass loss.
%We expect the \iso{13}C source to be negligible because in SAGB stars the region that can develop
%a \iso{13}C pocket is also tiny and probably negligible.
Further, SAGB stars have very small interpulse periods, more like 30--1000\,yr as 
opposed to 1-100\,kyr for AGB stars. Hence hundreds to thousands of pulses must be calculated.

\section{Main uncertainties}
We have indicated above some of the uncertainties in AGB evolution. We discuss these in more detail below. However,
the AGB is the last phase of evolution for these stars, and hence the models begin with the uncertainties 
already accumulated over all of the earlier phases. Thus we need to briefly review these if we are to 
provide a realistic estimate of the confidence we should place in the models.

\subsection{Extra-mixing on the first giant branch}
It is now well established that the predictions for
abundance changes resulting from first dredge-up are largely in agreement with observations.
It is also well established that there is a second mixing event that changes these
compositions, and it seems to begin at the position of the bump in the giant branch 
luminosity function (see [2] for a recent review, and [17] for a beautiful illustration using Li). 
The exact
mechanism for this extra-mixing is not known, and early investigations focussed on the obvious 
candidate of meridional circulation in rotating stars [18], but modern models suggests that this does
not match the observations [19,20]. Recent interest focusses on
thermohaline mixing following the
discovery of a molecular weight inversion that appears in RGB stars when they reach the bump [21]. 
Calculations show that this seems to match the observations reasonably well [22,23]. 
Debate exists concerning how to model this process, with 2D and 3D hydro calculations disagreeing 
with the typical 1D models used; see [24,25] and the extensive discussion in [2]. 
More work needs to be done before we can be confident in how to model this process, let alone verifying its
role in the observed abundance patterns.

\subsection{Extra-mixing on the asymptotic giant branch}
If some process causes extra-mixing on the RGB, does it also operate on the AGB? This has been postulated
by various authors as a possible explanation for O and Al isotope measurements in pre-solar grains [e.g.~26].
%26 Busso et al 2010
But there are also discrepancies with the C isotope ratios predicted for C stars.  
By the time the star has dredged-up sufficient \iso{12}C to produce C/O $> 1$ the ratio of \iso{12}C/\iso{13}C 
greatly exceeds the observed values [27]. There are also other problems that may be alleviated by some extra-mixing
%27 Karakas 2010
on the AGB. However, [28] showed that, at least for solar metallicities, the inclusion of the effects of 
extra-mixing on the RGB (usually ignored in the models) removed the problems on the AGB. The final word is yet to be
written.
%28 Karakas et al 2010

%\subsection{First dredge-up and the bump in the luminosity function}
%It seems that the limited data available for first dredge-up are in broad agreement with the
%models, as reviewed in [2]. However there is evidence that, at least for [Fe/H] around $-2$, the
%predictions of when the dredge-up begins are not in accord with observations [28]. %28=George paper
%This paper also shows that there is a discrepancy in the predicted magnitode of the bump in the luminosity function,
%as noted earlier by others [e.g.~29]. 
%29=Cassisi et al 2011
%The discrepancy seems to increase as [Fe/H] decreases, and can be alleviated by postulating 
%some overshoot inwards at the bottom of the convective envelope [28,30].
%30=DiCecco 

\subsection{Core helium burning}
The core helium-burning phase has a history of challenging our modelling skills. This is where semiconvection
was first recognised in the 70s [29,30] and later the core-breathing pulses added more unwelcome complications [e.g.~31].
It is now well documented that small variations in numerical details of the determination of the convective
boundaries can produce enormous differences in the size of the convective core, not to mention 
the semiconvection region, as discussed recently in [32]. The reality or otherwise of the core
breathing pulses remains in debate [e.g.~33] and recent work [34] has tried to use asteroseismology as a probe
of mixing in the cores of these stars. The problem is very difficult, with a core opacity source that is
higher (in the regions rich in C and O) than in the outer He-rich region. This drives
overshooting at the core edge. But the stellar conditions contrive to
produce a local minimum in the ratio of the adiabatic to radiative temperature gradients. When this minimum
reaches unity, the traditional value for convection according to the Schwarzschild criterion, then the
correct way to calculate the behaviour of the outer edge of the core is far from clear. Further, when the central
He content drops below about 0.1 [35] then the cubic dependence of the triple-$\alpha$ energy production 
on the He content means that small perturbations on the He mass fraction can produce large changes in the
energy output and hence drive larger convective cores -- these are breathing pulses.

Clearly the only way forward is multi-dimensional hydrodynamical simulations as discussed in [32]. Such
calculations are demanding because the simulation
must be performed for many turnover times. Indeed, determining the behaviour and timescale for semiconvection
is one of the hoped-for outcomes of such simulations.

\subsection{Convective boundaries}
Thankfully the timescale for convective mixing is usually much smaller than
the evolutionary timescale for the star.  Usually one uses a diffusion
equation to approximate mixing. Because mixing is usually very rapid, there is not
a lot of dependence on the diffusion coefficient, as long as it is sufficiently 
large to produce rapid mixing, although there are some notable exceptions, such as HBB.
As you can see from the previous discussion, the calculation of convective borders remains a 
serious problem for stellar models. Some overshoot must carry
material beyond the naive Schwarzschild border. This is simply the result of
conservation of momentum. The Schwarzschild criterion considers the buoyancy force, 
and places the border where that goes to zero. However material arrives at the neutral border 
with a finite momentum so it must penetrate the border -- but by how much? It has become common
to modify the diffusive mixing implementation by including an overshoot region where
the diffusion coefficient is chosen to reproduce an exponential decay in velocity beyond the formal 
border [36]. This is qualitatively fine, but as always there is a parameter (determining the
decay length) that is usually fixed by appealing to some observations. This procedure
produces the partial mixing required to produce a \iso{13}C pocket for $s$-process nucleosynthesis,
although of course the details depend on the overshoot procedure.

If one applies such a scheme to all convective borders, then interesting things happen. Of course, 
one must calibrate, or somehow choose, the decay length for each border, and there is no reason
to think that overshoot inwards (to higher density) has a similar decay length to 
overshoot outwards into less dense material. Nevertheless, when applied to the intershell convective
region, we find overshoot into the CO core with the result that the intershell
becomes enriched in C and O. Models including this effect give better
fits to observed abundances in H-deficient post-AGB stars [37] and possibly AGB stars as well [28].

\subsection{Convection theory}
Of course, the situation is even worse than described above. The Mixing-Length Theory (MLT) of 
convection has such a hold on stellar modelling that we forget that it is only one possible formulation.
Another that has made substantial contributions is the Full Spectrum of Turbulence  
theory [38,39]. These two theories produce significantly different results [40] and
it seems that this uncertainty is largely ignored in the literature. We note that  researchers
do continue to develop potential new convection models [e.g.~41] but these must be 
presented in a format that is easily implemented in an evolution code if they are to overcome the dominance of the MLT.

\subsection{Opacities for varying envelope compositions}
This is one area that has received a lot of attention recently, with the result that the current
situation is very satisfactory. Thermal pulses increase the C (and possibly O,
via overshoot, see above) content of the
stellar envelope, and HBB can burn this C (and O) into N. After H and He, these can be the next most abundant
species, and they are a very significant source of opacity. This is doubly so for stars with very low [Fe/H].
Such variations in the opacity have been ignored until recently, as tables for varying compositions were
not available. This is no longer the case, with the AESOPUS tool now providing opacities
for appropriate mixtures [42,43]. The increase in opacity when the C content increases has
also been shown to have a dramatic effect on the evolution of the AGB stars, increasing mass loss and 
hence terminating the evolution much sooner than in calculations that ignore this effect
[44,45].

\subsection{Envelope ejection?}
In 1986 Wood and Faulkner [46] found convergence problems in a late AGB model which they described as due to 
the disappearance of hydrostastic solutions to the stellar structure for large cores. Further work on this 
problem was performed by [47] who confirmed that a super-Eddington luminosity developed at the bottom
of the convective envelope. This was identified as being due to an opacity bump produced by Fe. An understanding
of the subsequent behaviour of the star will require a hydrodynamical study. Of course this may be important 
for understanding the formation of planetary nebulae.

\section{Final fate of AGB stars}
An AGB (or SAGB) star ends its life when mass loss removes the envelope. Usually this produces a CO white dwarf. However
the SAGB stars ignite core C and a number of outcomes become possible [3]. The C burning 
can ignite in a shell in the outer part of the CO core, and in many cases does not proceed further. 
This produces a CO(Ne) hybrid white dwarf.
If the C burning  proceeds to the core then we find the formation of an ONe white dwarf. 
The most massive SAGB stars experience ``dredge-out'' [48,49] where the convective C burning region meets with
the convective envelope. This is a computationally demanding phase of the evolution and subject to all the uncertainties 
associated with time-dependent mixing, which are amplified by simultaneous  rapid nuclear energy generation.

Following core He burning most SAGB stars have a core mass
that easily exceeds the Chandrasekhar mass. We expect such a star
to proceed through various nuclear burning stages and end life as a supernova.
But the occurrence
of second dredge-up in SAGB stars reduces the core mass
below the critical value. 
The fate of such a star depends on the competition
between core growth and mass loss. If the former dominates
and the core reaches the Chandrasekhar mass then an electron-capture supernova
will result. If mass loss terminates the evolution with the core mass below the critical
value then the star ends as an O–Ne white dwarf [3].

\section{Mass-loss}
From the viewpoint of stellar models, what is required is a formula that specifies the mass-loss 
rate (MLR) in terms of known quantities. This mass loss is of course assumed to be steady and
spherically symmetric, which we know is not always the case in reality. The mechanism believed to
drive mass loss in AGB stars is the pulsation enhanced dust-driven wind scenario, where grains 
are driven outward by the photon wind. These are collisionally linked to the gas, and hence
the gas is also removed. For stars with C/O $> 1$ we believe that amorphous C grains are involved.
For O-rich stars it is presumably Mg and Fe silicate grains that are implicated. The difficulty is
that these latter grains do not couple well with the gas unless they are very large [50]. It
is only recently that such large grains were indeed observed [51].

There are many MLR expressions in the literature. We discuss here only the most commonly used formulae.
The Reimers formula [52] 
was derived for giants and supergiants. Vassilliadis \& Wood [53] instead fit the MLR to 
the pulsation period for red giants and AGB stars. The MLR was bounded by 
the radiation limit, and for massive stars the superwind phase was delayed to ensure
that periods exceeding 500\,d were obtained. The Bl\"ocker formula [54] is one of many 
modifications to the Reimers rate, in this case motivated by dynamical pulsation models of Miras.
The more recent (2005) Schr\"oder \& Cuntz formula [55] is a physically motivated, semi-empirical 
modification of the Reimers formula.

As one may expect, the MLR has a potentially enormous effect on the star's evolution [6,40]
and nucleosynthesis [6,56]. Increasing the MLR removes the envelope more rapidly, thus terminating the
evolution and also terminating thermal pulses and all nucleosynthesis. The reultant yields are
very dependent on the MLR used (and the free parameters chosen for those formulae with such parameters).

Various authors have performed tests of the MLRs. Mostly these are crude sanity checks,
but some quantitative tests have also been performed. Schr\"oder \& Cuntz [57] looked at 
detailed models for some of the best studied galactic giants and supergiants, with
considerable success.  Another nice test was a critical examination of AGB luminosity 
functions [58] which again was very favourable to the Schr\"oder \& Cuntz MLR.

However, the Vassilliadis and Wood [53] formula also has been carefully tested.
Detailed evolution and pulsation models for thermaly pulsing stars in NGC419 and NGC1978
were compared with infrared data by [59] and an excellent agreement was found. 
The MLR in [53] also accurately predicted the magnitude of the tip of the AGB in these clusters.
Another quantitative test of this rate with AGB stars in the SMC was performed by [60] and again the MLR 
produced a successful quantitative comparison.

In conclusion, one should carefully choose the MLR to be used, depending on the phase of evolution 
being investigated. The Schr\"oder \& Cuntz  formula [55] seems suitable for giants and 
AGB stars, while the Vassilliadis and Wood [53] formula, tailored for AGB stars, does a very good 
job in that regime. 

%yyy
\section{Conclusions}
There remain many uncertainties in trying to model AGB and SAGB stars. As usual, these mostly centre
on convectioon and its various manifestations. Recent work has led to substantial improvements in
our understanding of mass loss and we now have formulae that
seem to be quantitatively reliable, at least in a global sense. The use of the Reimers formula 
for AGB stars is not recommended.

\ack JL thanks the organisers, and especially Sun Kwok and Linda Lee, for their 
hospitality and practical assistance. He also thanks Sun Kwok for 30 years of valued friendship.

\smallskip

%\bibitem{iopartnum} IOP Publishing is to grateful Mark A Caprio, Center for Theoretical Physics, Yale University, for permission to include the {\tt iopart-num} \BibTeX package (version 2.0, December 21, 2006) with  this documentation. Updates and new releases of {\tt iopart-num} can be found on \verb"www.ctan.org" (CTAN). 
%\end{thebibliography}

\end{document}